\documentclass{iau}
\usepackage{graphicx}

\title[Indigenizing the Drake Equation] %% give here short title %%
{Indigenizing the Drake Equation: how Indigenous methods can help us understand life in the Milky Way Galaxy}

\author[Hilding R.~Neilson]   %% give here short author list %%
{Hilding R.~Neilson$^1$
 }

\affiliation{$^1$Department of Physics \& Physical Oceanography, \\ Memorial University of Newfoundland \& Labrador \\ St.~John's, Newfoundland \& Labrador, Canada \\ email: {\tt hneilson@mun.ca} }

\pubyear{xxxx}
\volume{399}  %% insert here IAU Symposium No.
\pagerange{119--126}
\jname{Oxford XIII / Indigenous Astronomy in the Space Age}
\editors{D. Hamacher, \& J. Holbrook, eds.
}
\begin{document}

\maketitle

\begin{abstract}
The Drake Equation is a thought experiment whose purpose is to understand the ingredients necessary for life and advanced technological civilizations to exist on other worlds in our galaxy. However, beyond reflecting on life on Earth, we have no knowledge of many of these ingredients, such as the number of planets that have life, the number with intelligent life, the number with advanced civilizations, and the lifetimes of these civilizations. In this work, I will review the Drake Equation and the biases that scientists have traditionally had in discussing this equation and how it has led to the current searches of biological and technological signatures. I will discuss how the Drake Equation looks different if we consider it through the lens of Indigenous methods and sciences and how these methods would lead to a dramatically different view of life in our Galaxy.

\keywords{SETI, Indignenous Methods, Technosignatures}
\end{abstract}

\firstsection % if your document starts with a section,
\section{Introduction}
Perhaps one of the biggest questions for humanity is if there is life on other worlds, but it is only in the past seventy to eighty years has academic field taken it seriously.  Arguably some of the first searches for advanced civilizations was led by Dr. Frank Drake (\cite{Drake1962}) as he developed a thought experiment to predict the potential number of those civilizations that might be detectable from Earth in a search for radio emission.  Since then, the field a grown dramatically thanks to new ideas and initiatives such as the SETI institute.  

Today, astronomers search for  technosignatures to find the first evidence for advanced civilizations in the Galaxy. These searches continue to include searches for radio signals, but also optical searches (\cite{Acharyya2023, Zuckerman2023,Zuckerman2024}) for lasers and constructs such as Dyson Spheres.  These searches have built on the early work of Drake and, in particular, the Drake Equation.  In this work, I review the Drake Equation and consider the biases in the various terms of the Drake Equation that arise from the normative perspectives in western sciences.  I then discuss Indigenous methodologies for understanding nature and the universe around us and how these methodologies can allow us to reevaluate the Drake Equation to find different conclusions. 

\section{The Drake Equation}
The Drake Equation is an exercise for considering the various ingredients and probabilities for life to exist on another world, evolve to intelligence, and to become an advanced civilization that we on Earth could potentially detect.  The number of detectable civilizations in our galaxy is
\begin{equation}
    N = R_* f_Pn_ef_Lf_If_CL.
\end{equation}
The first term of the Drake Equation is the rate of star formation in the Galaxy and is $R_* \approx 2~M_\odot$~yr$^{-1}$ (\cite{Elia2022}), that is, we observe that there are about two Suns worth of mass are born every year. The second term $f_p$ is the fraction of stars that are born that will also form planets.  Thanks to the Kepler (\cite{Borucki2010}) and TESS surveys (\cite{Ricker2016}), about 20\% of stars appear to host planets (\cite{Zhu2018,Bryson2021, Boley2021, Ment2023}).  The variable $n_e$ is the average number of planets in a stellar system.  Our solar system hosts eight planets (or nine depending one's feelings about Pluto, or whether the Planet Nine exists (\cite{Brown2024}).  

The remaining terms centre on the formation and evolution of life.  The variable $f_L$ is the fraction of planets that can support the formation of life, while the variable $f_I$ is the fraction of systems that support life that will go on to evolve into intelligent life.  The variable $f_C$ is the fraction of systems that host intelligent life to evolve into advanced civilizations that build technologies that we might be able to detect. The final variable $L$ is the length of time that those civilizations are detectable, either because the advanced civilization moves on from that detectable technology or the civilization ceases to exist.  

This one equation has motivated numerous studies and searches, but, we continue to be limited by our ability to estimate many of the terms in the Drake Equation. For instance, the presence of life has only been confirmed by western science to be on the Earth and we have yet to detect life on any other planet or satellite in the Solar System. As such, we can state that $f_L >0$ but nothing more.  Likewise, there are similar issues with fractions of planets with intelligent life and with civilization.  As such, attempting to compute a potential number of civilizations is largely an academic exercise and the value of the Drake Equation is reminding us of our ignorance of those terms.

It should be noted, however, \cite[Hart (1975)]{Hart1975} argued that there can be no advanced civilizations in the Galaxy on the grounds that if they existed then at one of the civilizations would have visited the Earth at some point. This concept can be seen in the Drake Equation as the reduction of some of the fractions to a very small number.  This view of a vary small number of civilizations continues today. 

\section{Indigenous Methodologies}
The view of the Drake Equation comes from a Western or Eurocentric view and that view creates biases in our understanding of SETI and how the astronomy considers searches for life and so-called technosignatures. 

Indigenous methodologies offer a different perspective on SETi because it is built on different ways of knowing and different core beliefs. The key ``axioms'' of Indigenous methods and Western Science are compared in Table~\ref{tab1}, building on the works of \cite{Cajete2000, Battiste2013, Lipe2019} and others.

\begin{table}
  \begin{center}
  \caption{Overview of Indigenous ways of knowing versus Western Science.}
  \label{tab1}
 {\scriptsize
  \begin{tabular}{l|l}\hline 
{\bf Western Science} & {\bf Indigenous Methods}\\ 
\hline
 & What is above reflects below \\
Knowledge is objective & Knowledge is relational \\
Attempts to reduce to the smallest number of variables& Considers multiple variables concurrently    \\
 Knowledge is in disciplines &Knowledge is (w)holistic   \\
 Nature is hierarchic & Nature is sacred and familial \\
\end{tabular}}
\end{center}
\end{table}

It must be noted that there is really no such thing as pan-Indigenous knowledges or methods.  There can be Mi'kmaw knowledges, Salish knowledges, etc, but no pan-Indigenous methodologies.  Instead, we consider that many of the different Indigenous methodologies and ways of knowing have these axioms in common. 

The concepts in Table\ref{tab1} are significantly different from the methods of Western Science.  The first concept of \underline{What is above reflects below} is an acknowledgment that what is in the night sky is a reflection of the lands on which the peoples live . For instance, in Mi'kmak'i \footnote[1]{The homeland of the Mi'kmaw people in what is now called eastern Canada and northeastern US and is the First Nation of the author}, there is the constellation of Muin (Bear) and the Seven Bird Hunters and the story of that constellation is one that is specific to that land and the animals and seasons of that land.  

In Western Science, knowledge is objective (\cite{Shapin2018}).   If one wishes to conduct a scientific experiment then it should be reproducible and the in that reproduction another person should be able to obtain the same conclusion.  That is, the knowledge is independent of the person.  For many Indigenous peoples \underline{Knowledge is relational} and that can mean holding multiple truths that depend on one's experiences, even if they are in conflict (\cite{Deloria1, Smith2021}). Moreover, that relationality can require a responsibility for any new knowledge and how it is applied in the world.  In other words knowledge is not just something obtained but also something we have a responsibility about to use and share ethically.

In Western Science, we tend to try to understand one component of a phenomenon at any time, that is, we try to reduce the study to the smallest number of variables. For instance, we might study gravity in a physics lab by taking an object of known mass, dropping the object from various heights and measuring the time it takes for the ball to land.  In that experiment, we hold other variables constant and let the height be the only variable.  As such, we are trying to consider science only through the constituent parts. For many Indigenous knowledges, peoples {\underline{consider how multiple variables act concurrently}}.  For instance, in parts  of northern Canada, Caribou herds migrate over hundreds of kilometers in search of food and their migration path can change year-to-year.  The migration path is challenging to understand from a Western Science perspective because the migration depends on many variables and those variables can interact in different ways. However, Elders and Knowledge Keepers understand these migrations and know how to predict the migration path and know how to manipulate the migration path.  

The university, born in the so-called Enlightenment, is constructed in silos (\cite{Shapin2018}).  We have departments of physics, or math, or engineering, or social sciences, all tending to act independently of each other.  Their knowledges tend to be held in isolation and those times science in these fields cooperate and cross over, we celebrate the interdisciplinarity of the work as that is not the norm.  \underline{Indigenous knowledges are (w)holistic}, that is knowledge is not in silos.  Indigenous knowledges and stories of the stars are astronomy, but can also be stories of geography, biology, meteorology, etc.  These stories can be ethical and spiritual, hence the use of the word holistic. But, the word (w)holistic is used to emphasize that the knowledges span across all disciplines.

Finally, in Western Science, nature is considered a hierarchy with humans at the apex (\cite{Jensen2016, Kimmerer2013, Noon2022}).  Humans are more important than other mammals, plants, water, air and more.  This hierarchy also been used to justify racist ideas around white supremacy and intelligence (\cite{Saini2019}) But, today Western Science continues to employ that belief in hierarchy to support the current approach to manifest destiny in Outer Space (\cite{Trevino2020, Ellis2024}), the choice of locations of the next generation of telescopes (\cite{Neilson2019, Prescod2020}) and more.  For many, if not all, Indigenous Knowledges \underline{Nature is sacred and familial}.  This axiom goes beyond the idea that nature is a place of spirituality, but that animals, planets, water, etc, all have rights and ways of being that are equal to us.  Humanity exists not above nature, but with and because of Nature. As such many Indigenous peoples live in treaty with animals, plant and water and air and even outer space that expresses responsibilities for those non-human beings and ways of supporting them.

Altogether, this brief and incomplete summary of Indigenous Ways of Knowing and sciences show that they are a very different perspective on nature and phenomena than that of Western Science.  As such, applying Indigneous methods to big questions in sciences allows us to learn more about the Universe and life in it.  An analogy would be in astronomy where we have learned much about the Universe using optical observations. Yet, when we observe in radio or infrared or ultraviolet wavelengths or even detecting gravitational wave, the Universe around us looks very different.  Combining the different observations allows for a more complete and detailed vision of the Universe, even when the different observations disagree.

\section{Applying Indigenous Methods to the Drake Equation}
We can learn from these Indigenous methods and revisit the Drake Equation to explore some of the biases that arise from only considering Western Science perspectives. 

For the sake of this work, we will assume that the star formation rate $R_*$, the fraction of stars that will have planets $f_P$, and the average number of planets in a stellar system $n_e$ are unchanged if one considers Indigenous methodologies.  It is not obvious that this is reasonable as these terms in the Drake Equation are explicitly built on the assumption that life and civilizations can only arise around stars and on planets.  However, we will leave this question for future considerations. 

The fraction of planets that can host life is another challenging aspect to consider. The fraction is dependent on how life and the transition between life and non-life are defined.  NASA\footnote{https://astrobiology.nasa.gov/research/life-detection/about/} defines "Life is a self-sustaining chemical system capable of Darwinian evolution.''. This would contradict many Indigenous knowledges where the Sun, Moon, water, and stars can be life along with many things that Western Science would not consider life.  In many ways life can be defined by our relationships with it, not by the individual behavior of the life form.  It is not clear what this could mean for the Drake Equation, but the NASA definition is clearly limited by the Western lens. 

It is the remaining three terms of the Drake Equation that we will consider here.  
The first term, $f_I$, is the fraction of planets that will host intelligent life is arguably the most problematic term of the Drake Equation. \cite[Denning (2011)]{Denning2011} noted numerous issues with how the concept of intelligence is viewed with respect to SETI and particularly how only the view of intelligence as something that leads to advanced technology is important.  To build on this concern, intelligence in the Drake Equation is assumed to by humanity only, but only in the sense of humanity building technology.  This ignores the colonial and racist history of how intelligence is considered (\cite{Saini2019}), but also limits the potential for intelligence to only one species out of millions on Earth.  That definition follows from the hierarchical view of nature.  If we consider Nature as familial then if humans are intelligent then so much be our animal, planet and other non-human kin.  We may not understand their intelligences, but we cannot deny or ignore those intelligences. 

Based on the problematic definition of intelligence, we can simply ignore it going forward. As such for both Western and Indigenous perspectives we can use $f_I = 1$ or drop it without difficulty, though from a Western Science lens it could impact the number of civilizations.

The fraction of planets that will host civilizations, $f_C$, is also problematic. For the Drake Equation, the term refers to a global civilization built and centred on advanced technology that, for clarity, I will refer to as a technocivilization.  This ignores various historical and modern contexts for what is or is not a civilization and how civilizations can exist without being centred on the western model of technological advancement.  We can assume that our technocivilization is only about sixty years old, arising from the development of radio broadcasting and radar (\cite{Drake1962}).  But, this is not the only civilization on Earth.  It is beyond this work to debate what is civilization from the social science perspective, but it can be noted that Indigenous Nations are also civilizations that have existed since time immemorial.  That would include Mi'kmaw civilization, Inuit civilization, Salish civilization and more on Turtle Island, along with Indigenous Nations in Australia, South America, Africa and elsewhere. Hence,  from an Indigenous perspective we can assume that the value of $f_C$ is much greater than what we would assume from a Western perspective.

The final term of the Drake Equation is the detectable lifetime of a civilization, $L$. This term reflected the idea that technologies and their detectability are transient.  For instance, the original consideration of the Drake Equation was the detectability of radio signals based on the fact that human were leaking radio emissions into the cosmos as a byproduct of global communications. Over time our method of communication has changed and we do not leak radio emissions in the same way.  Today, our radio emissions are dominated by cellular technology and one might expect that emission to disappear as technology changes.  The other way to view the lifetime term is one of collapse, i.e., how long will a technocivilization exist before it collapses?  Historically on Earth that could have occurred and can still occur due to nuclear war, or human-induced climate change.  But, from other view of lifetime, one can estimate $L$ to be about the order-of-magnitude of one century old.

While the Drake Equation traditionally views $L$ as the time a technocivilization is detectable by humans at a fixed time, we can flip the definition to how long do civilizations exist.   Indigenous Nations have existed since time immemorial.  From a Western lens, Indigenous peoples have existed as civilizations for at least 60,000 years in Australia and more that 15,000 years in North America, if the most conservative western anthropological estimate is assumed.  Those timeframes are more than one hundred times longer than that of our Western technocivilization and still exist today.  

Based on these latter terms and the changes in how one might consider them assuming Indigenous methodologies, then from an Indigenous lens there would be many times more civilizations in the Galaxy than if we view the Drake Equation through a Western Lens.  From a Western lens, civilizations in our galaxy must be somewhat rare, while via an Indigenous lens life and civilizations would be abundant. These results are not in conflict, simply that the Western perspective of the Drake Equation is built to look for signatures of technology or technosignatures and those technosignatures tend to be considered as a reflection of our technocivilization and not other potential kinds of civilizations. But, by using Indigenous methods to reconsider the Drake Equation, we broaden the scope of the search for life in our galaxy.

%For instance, Indigenous Elder George Manuel wrote in The 4th World "TECHNOLOGIES ARE ONLY the tools through which we carry on our relationships with nature." (Manuel, 1974)

\section{Discussion}
By using Indigenous methods, it is possible to broaden the view of the Drake Equation and hence, consider how we search for technosignatures. This is done not by changing the scope of the equation but by considering the biases in how we use the Drake Equation. In particular, focusing on the Drake Equation through a lens of Western Science yields one kind of useful result while focusing on the equation through a lens of Indigenous methods suggests a very different results, one that has not been traditionally considered in the astronomy and SETI communities.  

Historically, technosignatures centred on artificial radio signals. Today, other technosignatures include: Dyson spheres, lasers, and more.  However, these technosignatures have one thing in common; they can only be observed as a significant deviation from a naturally occurring signal.  For instance, natural radio signals tend to be broad-lined with respect to wavelength, while artificial signals are narrow band.  Similarly, observations where authors have speculated about a Dyson sphere tend to be caused by varying eclipses of the host star.  Hence, we are not really looking for technosignatures but are looking for technoimpacts, or simply pollution.  We can carry the point further noting that one can consider pollution to be colonialism (\cite{Liboiron2021}), hence it is important to consider whether this current view of technosignatures may be colonial in nature.

\begin{figure}[t]
% \vspace*{-2.0 cm}
\begin{center}
 \includegraphics[width=5in]{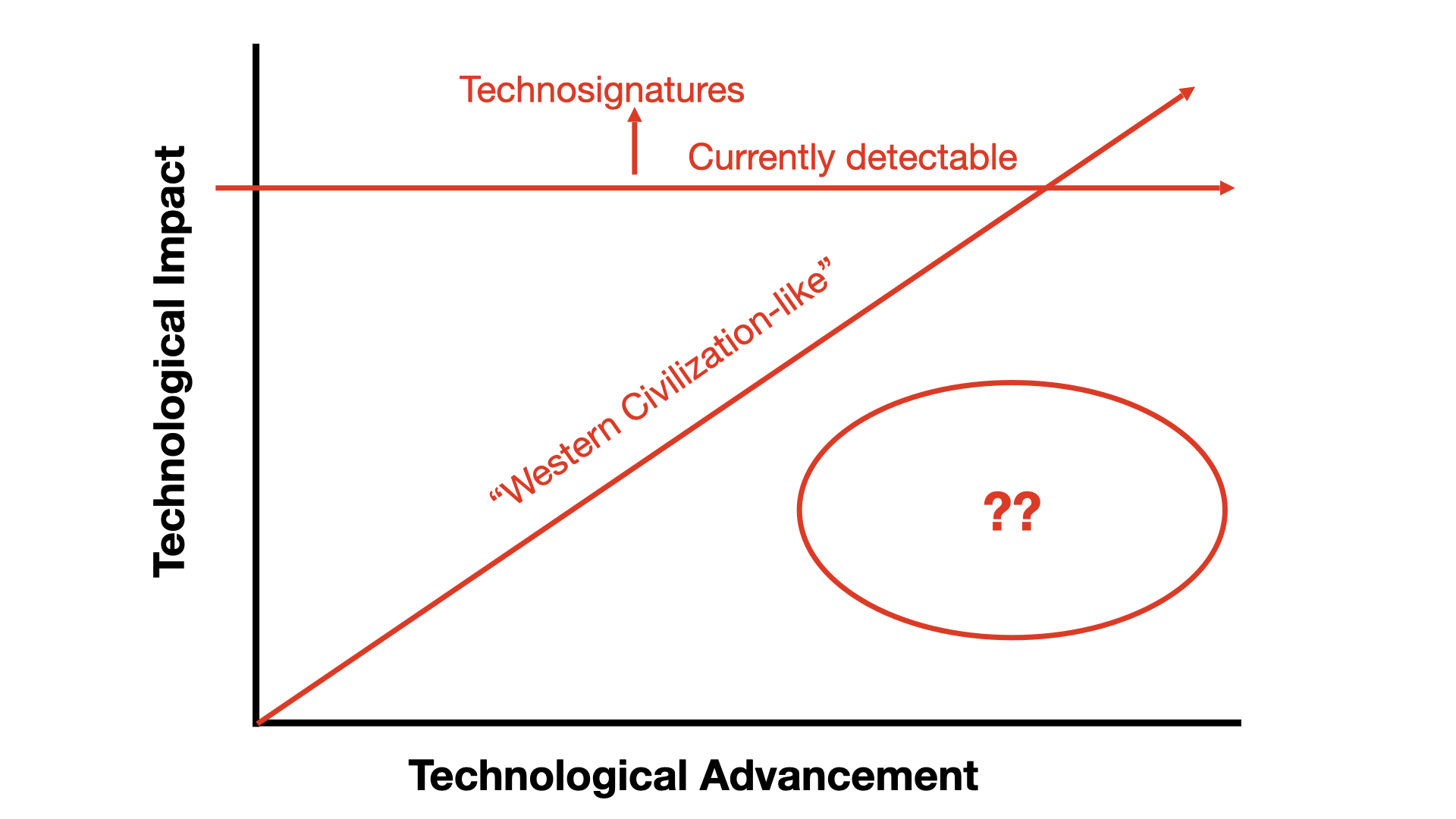} 
% \vspace*{-1.0 cm}
 \caption{Schematic diagram showing how civilizations can be varying levels of technological advance and technological impacts on nature. The diagonal red line represents western civilization while the horizontal line denotes the minimal technological impacts or technosignatures that can be detected. On the right where there is low impact but advanced technology would be group of civilizations that might be represented by Indigenous methods.  }
   \label{fig1}
\end{center}
\end{figure}

Indigenous Elder George Manuel wrote in The fourth world: an indian reality ``Technologies are only the tools through which we carry on our relationships with nature'' (\cite{Manuel2019}).  As such, Indigenous technologies must exist in support of nature and our responsibilities. Hence. the impacts of those technologies must be minimal. Therefore, one way to think about the search for technosignatures  is to consider that civilizations can have both advanced technologies by some definition as well as having varying technological impact. In Figure~\ref{fig1}, I show an illustration where societies can be considered with different levels of technological advancement and impact.  For simplicity, the western technocivilization is drawn as a diagonal line highlighting how most, if not all, of our technological advancements come with significant impacts on nature.  Our ability to detect technosignatures is constrained by the how technologies impact nature, suggesting this is largely independent of how advanced the technology is.  Since Indigenous technologies tend to be of lesser negative impact on nature then there is likely many civilizations with advanced technologies that we cannot detect because the impact is too small. 
Furthermore, one can flip the perspective of this cartoon and ask whether extraterrestrial civilizations can detect the Earth and its technological impacts.  In this situation, civilizations with advanced technologies would likely be able to detect other civilizations with less technological impact than what humans might be able to detect. That is, there are potentially advanced civilizations that we cannot detect but they could detect us.  How would those civilizations perceive our Western-like civilization?  Perhaps, they would see us as colonial and this would be another potential resolution to the Fermi Paradox (e.g., \cite{Webb2015}). That is, it is possible that these extraterrestrial civilizations with technologies that support and do not harm nature would be anti-colonial.

\section{Summary}
In this work, I illustrate how we can apply Indigenous methods to explore the Drake Equation and learn about the potential for civilizations in the Galaxy.  By using these methods, we would predict that there are many more civilizations in the Galaxy than if we simply apply the traditional western scientific methods.  As such Indigenous methods and western methods offer two very different perspectives of life in the Galaxy. 

When this result is considered via how humans tend to search for technosignatures, I show that the traditional view of technosignatures would omit the presence of many civilizations because those civilizations have advanced technologies but lesser technological impacts on the environments.  As such, current searches tend to focus only on a small fraction of civilizations in the Galaxy.  

The two views of life in the Galaxy can be considered together using the concept of Two-Eyed Seeing, or Etuaptmumk (\cite{Bartlett2012}) where bringing Indigenous knowledges and western science methods together we can develop a deeper understanding.  In this way, the two methods together show that we need to reconsider the concepts of technosignatures in the search for civilizations and develop searches that consider non-western perspectives.

\end{document}